\documentstyle [preprint,float,aps,psfig]{revtex}
\tightenlines

\begin{document}

\title{Measurement of the Cross Section Asymmetry \\
of the Reaction  $\gamma p\to \pi^{\circ} p$ in 
the Resonance Energy \\
Region $E_{\gamma} = 0.5 - 1.1\:$GeV}

\author{F. V.~Adamian, 
A. Yu.~Buniatian,
G. S.~Frangulian\thanks{On leave from Yerevan Physics Institute},
P. I.~Galumian\thanks{On leave from Yerevan Physics Institute},
V. H.~Grabski, 
A. V.~Hairapetian, 
H. H.~Hakopian, 
V. K.~Hoktanian, 
J. V.~Manukian, 
A. M.~Sirunian, 
A. H.~Vartapetian, 
H. H.~Vartapetian, and
V. G.~Volchinsky}
\address{Yerevan Physics Institute, 2 Alikhanian Brothers St.
Yerevan, 375036 Armenia}

\author{R. A.~Arndt,
I. I.~Strakovsky, and
R. L.~Workman}
\address{Center for Nuclear Studies and Department of Physics\\
The George Washington University, Washington, DC 20052, USA}

\draft
\date{\today}
\maketitle

\begin{abstract}
The cross section asymmetry $\Sigma$ has been measured 
for the photoproduction of $\pi^{\circ}$-mesons off 
protons, using polarized photons in the energy range 
$E_{\gamma} = 0.5 - 1.1\,$GeV.  The CM angular coverage 
is $\theta_{\pi}^{*} = 85^{\circ} - 125^{\circ}$ with 
energy and angle steps of 25\,MeV and $5^{\circ}$, 
respectively.  The obtained $\Sigma$ data, which cover 
the second and third resonance regions, are compared 
with existing experimental data and recent 
phenomenological analyses.  The influence of these 
measurements on such analyses is also considered. \\
\end{abstract}


\narrowtext
\section{Introduction} 
\label{sec:intro}

Single-pion photoproduction has been used 
extensively to explore the electromagnetic 
properties of nucleon resonances, and most 
determinations of the $N \gamma$ resonance 
couplings \cite{pdg} have been obtained through 
multipole analyses of this reaction.  The study 
of $N^*$ properties, in general, has enjoyed a 
resurgence, driven by the growth of new 
facilities worldwide.  Many precise new 
measurements have focused on the photoproduction 
of pions and other pseudoscalar mesons, in the 
hope that they would reveal states not seen 
previously in the photoproduction and elastic 
scattering of pions.  Associated studies have 
suggested both new states and couplings to 
existing states which contradict those found in 
analyses of the full pion production database.

These results demonstrate the influence of 
measurements sensitive to new quantities and 
suggest that older analyses may have been based 
on insufficient or flawed data sets.  A survey 
of existing data in the 1~GeV region shows most 
polarization measurements to have only one or 
two angular points at a given beam energy, 
often with rather large uncertainties.  This 
allows a great deal of freedom in multipole 
analyses, and is clearly insufficient to pick 
out more than the strongest resonance signals.

One quantity which has consistently influenced 
these new analyses is the beam polarization 
quantity $\Sigma$.  Examples include the E2/M1 
ratio \cite{e2m1}, where new measurements of 
the differential cross section and $\Sigma$ 
changed this ratio by nearly a factor of two.  
Measurements of $\Sigma$ in $\eta$ 
photoproduction were shown 
\cite{etaphoto1,etaphoto2} to be sensitive to 
the $A_{3/2} / A_{1/2}$ ratio of photo-decay 
amplitudes for the $D_{13}(1520)$, implying a 
ratio very different from the pion 
photoproduction value.  The behavior of 
$\Sigma$ in kaon photoproduction has also 
proved \cite{kaonphoto} to be crucial in 
determining the character of a bump seen in 
precise new total cross section measurements.  
In $\pi^+ n$ photoproduction, results for 
$\Sigma$ from GRAAL have been used to argue 
for a change in the $A_{1/2}$ amplitude for 
the $S_{11}(1650)$ \cite{GRAAL}.

In this experiment, we have performed systematic
measurements of the cross section asymmetry 
$\Sigma$ for the reaction $\gamma p\to\pi
^{\circ} p$ \cite{vartap} simultaneously with 
measurements of Compton scattering $\gamma p
\to\gamma p$ \cite{vartap,adam}, over the 
energy range $E_{\gamma}$ = 0.5 - 1.1\,GeV 
and at pion CM angles $\theta_{\pi}^{*} = 
85^{\circ} - 125^{\circ}$.  The experimental 
data (158 data points) from $\pi^{\circ}$-meson 
photoproduction constitute the first systematic 
and high statistics measurement of $\Sigma$ for 
this reaction channel over the present kinematic 
region.  Preliminary results of the measurements 
have been presented in \cite{pi0}.  

In Sections~2 and 3, we describe the methods 
employed in this experiment, with special 
emphasis on the control of systematic errors.  
In Section~4, we describe how these measurements, 
and the recent measurements from GRAAL 
\cite{graalnew}, have effected the GW multipole 
analyses.  Comparisons to the Mainz fits 
\cite{mainz} are also made.  Finally, in Section~5, 
we summarize our results.

\section{Experimental Setup}
\label{sec:ExpSet}

The experiment was carried out on the linearly 
polarized photon beam of the 4.5\,GeV Yerevan 
Synchrotron.  The same setup has been used for 
the parallel study of two reactions: Compton 
scattering and photoproduction of $\pi
^{\circ}$-meson off protons.  The experimental 
setup is shown in Fig.\,\ref{f1}. 

The linearly polarized photon beam is generated 
by coherent bremsstrahlung (CB) of 3.0--3.5\,GeV 
electrons on the internal $100\,\mu$m diamond 
crystal target.  The beam has 2--3\,msec long 
pulses with a 50\,Hz repetition rate and an 
intensity of about $5\times10^9$ eq.~photons/sec.  
The beam, shaped by a system of collimators and 
sweeping magnets, has a 10$\times$10\,mm size at 
the position (${\rm H_2}$) of the liquid hydrogen 
target (${\rm H_2}$) (9\,cm in length).  It is 
transported in a vacuum pipe to the Wilson-type 
quantameter (Q).  The energy spectrum is measured 
and controlled by a 30-channel pair spectrometer 
PS-30 (energy resolution 1--2\%) \cite{adam2}.  
The whole range of bremsstrahlung energies are 
covered by 5 scans with respective current 
settings on the PSM analyzing magnet.  The full 
energy spectra of bremsstrahlung on amorphous 
and crystal targets, measured by the pair 
spectrometer PS-30, are presented in Fig.\,
\ref{f2}.

The recoil protons are detected by the magnetic 
spectrometer (MS) consisting of a doublet of 
quadrupole lenses (${\rm L_1}$, ${\rm L_2}$), a 
bending magnet (BM), a telescope of four 
scintillation trigger counters (${\rm S_1-S_4}$) 
and a system of coordinate detectors, including 
a scintillation hodoscope (${\rm H_p}$) and 
seven multiwire proportional chambers (${\rm MWPC
_{xy}}$), allowing the reconstruction of momentum, 
azimuthal and polar angles of the outgoing proton.  
Time-of-flight measurements were carried out 
between ${\rm S_1}$ and ${\rm S_4}$ counters on a 
flight base of 9~m for particle identification 
(p,\,$\pi$).  The MS covered a solid angle of 
$\Delta\Omega\approx 3.5\,msr$, and its angular 
and momentum resolutions were $\sigma_{\theta}
\approx 0.3^{\circ}$, $\sigma_{\phi}\approx 
0.2^{\circ}$ and $\sigma_P/P\approx 1\%$, 
respectively.  The incident photon energy and the 
CM scattering angle were reconstructed on average 
with an accuracy of $\sigma_{E_{\gamma}}/E_{\gamma}
\approx 1\%$ and $\sigma_{\theta^{*}} \approx 
0.6^{\circ}$, with corresponding acceptances of 
$\approx 18\%$ and $\approx 7.5^{\circ}$, 
respectively.

The $\pi^{\circ}$ decay photons were detected by 
the ${\rm \check{C}}$erenkov spectrometer (${\rm 
\check{C}_s}$) consisting of a veto counter (V), 
a lead converter (Pb), scintillation hodoscope 
($\rm H_{xy}$) for $x$ and $y$ coordinate analysis, 
trigger counter (T), and a lead glass ${\rm 
\check{C}}$erenkov counter of full absorption ($\rm 
\check{C}$).  The energy resolution of the ${\rm 
\check{C}}$erenkov counter could be parameterized 
as $\sigma(E)/E = 0.08/\sqrt{E}$ without converter 
and $\sigma(E)/E = 0.1/\sqrt{E}$ with converter (E 
being the photon energy in GeV).

The kinematics of the analyzed process was completely 
determined by defining the kinematic parameters of 
the proton in the magnetic spectrometer.  For 
identification of the process $\gamma p\to\pi^{\circ} 
p$, the protons in the magnetic spectrometer were 
detected in coincidence with the photon detection 
branch. The coordinate detectors of the ${\rm 
\check{C}}$erenkov spectrometer were in addition used 
for correlation analysis between two detecting 
branches to identify the low rate yield of the 
Compton scattering process \cite{vartap,adam}. 

An independent experimental study was carried 
out to check the reconstruction accuracy in the 
magnetic spectrometer (MS) \cite{vartap}.  Monte 
Carlo calculations, intended to estimate the 
influence of different possible factors on the 
reconstruction precision, indicate that 
inaccuracies in setting the positions, angles and 
fields of magnetic elements, poor quality of 
track reconstruction in ${\rm MWPC _{xy}}$, 
additional multiple scattering in materials 
not taken into account, influence equally the 
reconstruction accuracy of both the interaction 
vertex and the scattering angles in MS.  Therefore, 
an experimental study of the vertex position 
reconstruction accuracy was defined as an effective 
test of the characteristics of the MS, and was 
easily carried out by means of a point-like target.  
Good agreement between the experimental results for 
vertex reconstruction and Monte Carlo calculations 
(Fig.\,\ref{f3}) provides reliable support for the 
performance of our proton detection branch within 
designed precisions.

Finally, the possibility to treat Compton scattering 
data \cite{vartap,adam}, despite significant 
difficulties caused by the physical background process 
of $\pi^{\circ}$ photoproduction with approximately 
two orders of magnitude higher cross section, and with 
practically the same kinematic parameters, also points 
to a reliable operation of the detection system within 
designed accuracies.

\section{Experiment and Data Analysis}
\label{sec:ExpDA}

During the measurements, several factors were considered 
to minimize the influence of systematic uncertainties.  
A particular emphasis has been placed on maintaining the 
stability of the coherent peak position.   In the case 
of its deviation, the event registration was paused, 
until respective correction procedure was made.  Such 
checks of the coherent peak were carried out every 
40--50\,sec.  In Fig.\,\ref{f4}, the measured peak 
region is presented, superimposed on the full coherent 
bremsstrahlung spectrum, for the cases with correct and 
distorted peak position.  A possible influence on the 
asymmetry results from instabilities in subsystems was 
compensated by periodical alternating the orientation 
of the photon beam polarization during the measurement 
of the same kinematic point.  Also, during data 
acquisition the trigger rates, efficiencies of the 
coordinate detectors, distributions of kinematic 
variables were controlled.  Hardware and software of 
all these control procedures was realized through 
the CAMAC standard.  Fiber optic lines were used to 
manage the current of the pair spectrometer magnet (PSM) 
and the electronics of the MS trigger, placed immediately 
on the MS.

During the off-line analysis, prior to the reconstruction 
step, the time-of-flight measurements between ${\rm S_1}$ 
and ${\rm S_4}$ were analyzed, and the events with a 
proton candidate in MS were selected (Fig.\,\ref{f5}).  A 
soft cut was made, mainly at low energies, on the energy 
response of the ${\rm \check{C}}$erenkov detector, to 
reduce the contribution of accidental coincidences and 
multiparticle events resulting from the energetic part of 
the coherent bremsstrahlung spectrum \cite{vartap}.  
Then, the selected events were sent for reconstruction of 
the kinematic parameters of proton and the kinematics of 
the event. 

Within further selection, the interaction point 
in the target was reconstructed and events 
falling out of the target limits by more than 
one standard deviation are cut. 

The yield of the $\gamma p\to\pi^{\circ} p$ 
reaction was defined from the number of selected 
events in the time spectrum of start-stop 
measurements between two detection branches 
(Fig.\,\ref{f6}), by fitting and subtracting the 
respective number of background events in the 
peak region.

Several corrections were applied to compensate 
for the contribution of various factors.  The 
results from the analysis of Compton scattering 
process were used to correct for the influence 
of that source.   Further corrections were 
applied, compensating for the rate reductions 
due to the dead time of data acquisition and 
efficiency of different subdetectors.  The 
maximum trigger rates was about 5\,event/sec, 
and the dead time did not exceed 10\%.  Another 
insignificant contribution, resulting from the 
empty target cell, was measured separately and 
amounted to $\sim$1\%.

The background $p\gamma-$ rate is mainly 
dominated with double pion production processes 
generated by the high energy tail of CB spectrum. 
This background contribution  has been determined 
in additional measurements, such as:
\begin{itemize}
\item
$p\gamma-$rate dependence on the coherent peak
energy of the CB spectrum at fixed kinematics
of the setup, allowing the significant variation 
in  the relative yield of single and multiple 
pion production processes.
\item
The $p\gamma-$rate dependence on the $\pi 
^{\circ}-$photon decay angle.
\end{itemize}

The data  obtained  were then compared with 
results of Monte Carlo calculations.  The 
contribution of background processes was 
estimated to be less than 5\%.

Calculations of the polarization of the CB 
spectra were realized by a solution of integral 
equations for intensity and polarization of the 
coherent spectra \cite{hrach}.  In the energy 
range of the measurements, the photon beam, 
polarization ranged from 50\% to 70\%.
  
The cross section asymmetry $\Sigma$ of the 
reaction $\gamma p\to\pi^{\circ}p$ is 
defined as:
\begin{eqnarray}
\Sigma = \frac{1}{P_{\gamma}}\frac{C_{\perp} - 
C_{\parallel}}
                                  {C_{\perp} + 
C_{\parallel}}\;\;\; ,
\label{1}\end{eqnarray}
where $P_{\gamma}$ is the polarization of photon 
beam and $C_{\perp,\parallel}$ are the normalized 
yields for the photons polarized perpendicular 
and parallel to the reaction plane.  

The systematic errors of the measurements were 
dominated mainly by the accuracy of the linear 
polarization calculations.  The uncertainties 
in $\sigma P_{\gamma}/P_{\gamma}$ have been 
evaluated to be within $\approx 2.3\%$ in full 
energy range and are included in the errors of 
the experimental data.  The possible shift in 
the reconstructed photon energies was 
calculated from the uncertainty in the MS 
optics and did not exceed 0.4\%.  Finally, 
overall systematic uncertainty for each energy 
set did not exceed 3\%.

It is also important to note, that most 
kinematic bins were formed from at least two 
independent data samples collected at different 
kinematic settings of the detector.  The final 
asymmetry was calculated from the corresponding 
results in such overlapping bins.  On the other 
hand, comparison of the results on such 
equivalent bins, produced from different event 
samples, with different acceptance conditions 
and polarized incident photons from different 
parts of the CB spectrum, provided a good check 
for systematic uncertainties in the obtained 
results.

\section{Results and Discussion}
\label{sec:ResDis}

Our data for the asymmetry $\Sigma$, in the 
kinematic range $E_{\gamma}$ = 0.5 $-$ 1.1\,GeV 
and $\theta_{\pi}^{*} = 85^{\circ} - 125
^{\circ}$, are presented in Table\,1.  Some of 
the measured energy and angular distributions 
are shown in Figs.\,\ref{f7} $-$ \ref{f9} 
together with the experimental data of other 
groups \cite{zdar,alest,knies,ganen,yerold} 
and predictions of different phenomenological 
analyses \cite{maid2,arndt}. 
  
The most systematic measurements of $\Sigma$ 
for $\gamma p\to\pi^{\circ} p$, over the 
resonance region, exist for the angle $\theta
_{\pi}^* = 90^{\circ}$ (Fig.\,\ref{f7}a).  
From Fig.\,\ref{f7}a, one can see, as a whole, 
a good agreement between the values of prior 
measurements and our present results.  Near 
900~MeV, our data disagree with data from MIT 
\cite{alest} and some early SLAC \cite{zdar} 
measurements, but are in good agreement with 
later measurements from another SLAC group 
\cite{knies}.  Our points are also consistent 
with the results of this group for the angle 
\mbox{$\theta_{\pi}^* = 110^{\circ}$} (Fig.\,
\ref{f8}b).  From Fig.\,\ref{f9}, it is seen 
that our results for the angular distributions 
at $E_{\gamma}$ = 700, 750, and 800~MeV are 
in agreement with the existing data as well.   
 
In Figs.\,\ref{f7} $-$ \ref{f9}, the 
experimental data are also compared with the 
predictions of phenomenological analyses from 
the Mainz \cite{maid2} and GW \cite{arndt} 
groups.  While both analyses reproduce the 
qualitative behavior of $\Sigma$, a shape 
discrepancy is noticeable in the GW fit.  The 
main difference between multipole analyses 
and experiment is observed above $E_{\gamma}$ 
= 700~MeV where there have been few previous 
measurements.  Inclusion of our data in the 
GW fit results in an improved description, but 
shape differences remain.  The Mainz fit 
appears to have a shape more consistent with 
the data.  (It should be noted that some of 
the older fits, in particular the fit of Ref.
~\cite{barb}, predict a shape consistent 
with the Mainz result.)

It is interesting to compare these results to 
those recently published by the GRAAL 
collaboration \cite{GRAAL}.  In that work, a 
similar comparison was made for $\Sigma$ 
measurements in the reaction $\gamma p\to \pi^+ 
n$.  There a deviation from GW predictions was 
noted in the 800 $-$ 1000~MeV range, at 
backward angles.  It was suggested that this 
discrepancy could be removed by a change in 
the N(1650) photo-decay amplitude ($A_{1/2}$).  
We have examined the GRAAL angular distribution 
at 950~MeV ($W_{CM}$ = 1630~MeV) and agree that 
a change (reduction) in the $E_{0+}^{1/2}$ 
multipole can account for the shape difference.  
Here too, for neutral pion production and 
energies near 900~MeV, a reduction of this 
multipole in FA00 by about 20\% (in modulus) 
results in an improved description.  A 
reduction of the same amount is found in 
comparing the energy-dependent and 
single-energy solutions.  The effect is 
displayed in Fig.\,\ref{f10}.  Note that this 
modification to the GW fit, while improving 
the agreement with data, actually worsens the 
agreement with the Mainz prediction at the 
most backward angles.  

\section{Conclusion}
\label{sec:conc}

The obtained data on cross section asymmetry 
($\Sigma$) are in good agreement with existing 
experimental data and in at least qualitative 
agreement with phenomenological predictions 
\cite{maid2,arndt,barb}.  We have found 
evidence in support of the GRAAL claim that 
some discrepancies with previous GW analyses 
could be linked to the $E_{0+}^{1/2}$ multipole 
and possibly the N(1650) resonance contribution.  
It will be useful to have a Mainz fit including 
these data in order to isolate differences with 
the GW fits.  A set of fits employing identical 
data sets would also be useful in determining 
whether differences are due to the chosen 
formalism or data constraints.  Work along this 
line is in progress \cite{tiator}.

In summary, results of present experiment on 
$\Sigma$ asymmetry have significantly improved 
the existing data base in the second and third 
resonance regions.  Forthcoming measurements of 
cross section and polarization observables in 
the photoproduction of mesons (JLab 
\cite{cebaf1,cebaf2}, GRAAL \cite{graalnew}), 
may lead to more successful determinations of 
the underlying scattering amplitudes and a more 
precise determination of resonance 
photocouplings.
                         
\acknowledgments
The YERPHI group is indebted to the synchrotron 
staff and cryogenic service for reliable 
operation during the experiment.  This work was 
supported in part by the Armenian Ministry of 
Science (Grant-933) and the U.~S. Department 
of Energy Grant DE--FG02--99ER41110.  The GW 
group gratefully acknowledges a contract from 
Jefferson Lab under which this work was done.  
Jefferson Lab is operated by the Southeastern 
Universities Research Association under the 
U.~S.~Department of Energy Contract 
DE--AC05--84ER40150.

\eject

\newpage                

\setcounter{table}{0}
\begin{table}[htb] \centering                                                   
\begin{tabular} {|c|c|c||c|c|c||c|c|c|}                                         
\hline                                                                          
  \multicolumn{3}{|c||}{$\theta^{*} = 85^{\circ}\;\pm\; 2.5^{\circ}$} &           
  \multicolumn{3}{|c||}{$\theta^{*} = 90^{\circ}\;\pm\; 2.5^{\circ}$} &           
  \multicolumn{3}{|c|} {$\theta^{*} = 95^{\circ}\;\pm\; 2.5^{\circ}$} \\           
\hline                                                                          
  $E_{\gamma}\:$(MeV) & $\Sigma$ & $\sigma_{\Sigma}$ &                          
  $E_{\gamma}\:$(MeV) & $\Sigma$ & $\sigma_{\Sigma}$ &                          
  $E_{\gamma}\:$(MeV) & $\Sigma$ & $\sigma_{\Sigma}$ \\                         
\hline                                                                          
 500 $\pm$ 12.5 & 0.564 & 0.081 & 500 $\pm$ 12.5 & 0.643 & 0.060 &             
                                  500 $\pm$ 12.5 & 0.643 & 0.060 \\            
 525 &  0.683 & 0.052 &  525 &  0.636 & 0.039 &  525 &  0.617 & 0.039 \\            
 550 &  0.572 & 0.039 &  550 &  0.600 & 0.031 &  550 &  0.636 & 0.031 \\            
 575 &  0.638 & 0.034 &  575 &  0.640 & 0.028 &  575 &  0.627 & 0.029 \\            
 600 &  0.686 & 0.032 &  600 &  0.681 & 0.027 &  600 &  0.693 & 0.030 \\            
 625 &  0.720 & 0.031 &  625 &  0.718 & 0.027 &  625 &  0.715 & 0.033 \\            
 650 &  0.764 & 0.031 &  650 &  0.776 & 0.028 &  650 &  0.702 & 0.041 \\            
 675 &  0.828 & 0.031 &  675 &  0.802 & 0.030 &  675 &  0.756 & 0.063 \\            
 700 &  0.864 & 0.032 &  700 &  0.845 & 0.035 &  700 &  0.878 & 0.032 \\            
 725 &  0.876 & 0.033 &  725 &  0.843 & 0.025 &  725 &  0.864 & 0.027 \\            
 750 &  0.854 & 0.032 &  750 &  0.842 & 0.025 &  750 &  0.823 & 0.025 \\            
 775 &  0.812 & 0.031 &  775 &  0.746 & 0.022 &  775 &  0.711 & 0.023 \\            
 800 &  0.747 & 0.030 &  800 &  0.679 & 0.021 &  800 &  0.640 & 0.023 \\            
 825 &  0.675 & 0.031 &  825 &  0.671 & 0.021 &  825 &  0.601 & 0.025 \\            
 850 &  0.577 & 0.024 &  850 &  0.541 & 0.022 &  850 &  0.497 & 0.026 \\            
 875 &  0.515 & 0.026 &  875 &  0.475 & 0.024 &  875 &  0.394 & 0.025 \\            
 900 &  0.397 & 0.028 &  900 &  0.353 & 0.025 &  900 &  0.304 & 0.026 \\            
 925 &  0.356 & 0.030 &  925 &  0.223 & 0.023 &  925 &  0.144 & 0.028 \\            
 950 &  0.196 & 0.031 &  950 &  0.076 & 0.024 &  950 & -0.045 & 0.031 \\           
 975 &  0.168 & 0.032 &  975 & -0.002 & 0.026 &  975 & -0.108 & 0.047 \\         
1000 &  0.105 & 0.034 & 1000 & -0.054 & 0.028 & 1000 & -0.144 & 0.084 \\        
1025 &  0.097 & 0.032 & 1025 & -0.108 & 0.039 &      &        &       \\          
1050 &  0.103 & 0.042 & 1050 & -0.159 & 0.064 &      &        &       \\          
1075 & -0.032 & 0.061 &      &        &       &      &        &       \\            
\hline                                                                          
\end{tabular}                                                                   
\caption{\it The experimental results on $\Sigma$ asymmetry of                  
             $\pi^{\circ}$ photoproduction.}                        
\end{table}                                                                   
\setcounter{table}{0}
\begin{table}[htb] \centering                                                   
\begin{tabular} {|c|c|c||c|c|c||c|c|c|}                                         
\hline                                                                          
  \multicolumn{3}{|c||}{$\theta^{*}=100^{\circ}\;\pm\; 2.5^{\circ}$} &          
  \multicolumn{3}{|c||}{$\theta^{*}=105^{\circ}\;\pm\; 2.5^{\circ}$} &          
  \multicolumn{3}{|c|} {$\theta^{*}=110^{\circ}\;\pm\; 2.5^{\circ}$} \\          
\hline                                                                          
  $E_{\gamma}\:$(MeV) & $\Sigma$ & $\sigma_{\Sigma}$ &                          
  $E_{\gamma}\:$(MeV) & $\Sigma$ & $\sigma_{\Sigma}$ &                          
  $E_{\gamma}\:$(MeV) & $\Sigma$ & $\sigma_{\Sigma}$ \\                         
\hline                                                                          
  700 $\pm$ 12.5 & 0.867 & 0.029 & 700 $\pm$ 12.5 & 0.851 & 0.029 &             
                                   700 $\pm$ 12.5 & 0.803 & 0.027 \\            
  725 & 0.824 & 0.029 & 725 & 0.794 & 0.027 & 725 & 0.778 & 0.026 \\            
  750 & 0.772 & 0.027 & 750 & 0.720 & 0.024 & 750 & 0.708 & 0.024 \\            
  775 & 0.724 & 0.026 & 775 & 0.692 & 0.024 & 775 & 0.649 & 0.024 \\            
  800 & 0.648 & 0.025 & 800 & 0.645 & 0.023 & 800 & 0.587 & 0.028 \\            
  825 & 0.627 & 0.026 & 825 & 0.585 & 0.025 & 825 & 0.478 & 0.038 \\            
  850 & 0.595 & 0.029 & 850 & 0.524 & 0.032 & 850 & 0.360 & 0.085 \\            
  875 & 0.482 & 0.038 & 875 & 0.382 & 0.052 &     &       &       \\              
  900 & 0.211 & 0.058 & 900 & 0.204 & 0.089 &     &       &       \\              
\hline                                                                          
\end{tabular}                                                                   
\caption{\it (continued).}                  
\end{table}                                                                   
\setcounter{table}{0}
\begin{table}[htb] \centering                                                   
\begin{tabular} {|c|c|c||c|c|c||c|c|c|}                                         
\hline                                                                          
  \multicolumn{3}{|c||}{$\theta^{*}=115^{\circ}\;\pm\; 2.5^{\circ}$} &          
  \multicolumn{3}{|c||}{$\theta^{*}=120^{\circ}\;\pm\; 2.5^{\circ}$} &          
  \multicolumn{3}{|c|} {$\theta^{*}=125^{\circ}\;\pm\; 2.5^{\circ}$} \\          
\hline                                                                          
  $E_{\gamma}\:$(MeV) & $\Sigma$ & $\sigma_{\Sigma}$ &                          
  $E_{\gamma}\:$(MeV) & $\Sigma$ & $\sigma_{\Sigma}$ &                          
  $E_{\gamma}\:$(MeV) & $\Sigma$ & $\sigma_{\Sigma}$ \\                         
\hline                                                                          
 500 $\pm$ 12.5 & 0.637 & 0.071 & 500 $\pm$ 12.5 & 0.717 & 0.070 &             
                                                 &       &       \\            
 525 &  0.728   & 0.049 &  525  &  0.677 & 0.043 &                                   
                                     525 $\pm$ 12.5 & 0.739  & 0.051 \\            
 550 &  0.750 & 0.040 &  550 &  0.724 & 0.033 & 550 &  0.744 & 0.034 \\            
 575 &  0.792 & 0.036 &  575 &  0.686 & 0.028 & 575 &  0.718 & 0.028 \\            
 600 &  0.742 & 0.033 &  600 &  0.718 & 0.029 & 600 &  0.700 & 0.027 \\            
 625 &  0.744 & 0.032 &  625 &  0.746 & 0.029 & 625 &  0.649 & 0.032 \\            
 650 &  0.770 & 0.032 &  650 &  0.792 & 0.027 & 650 &  0.744 & 0.052 \\            
 675 &  0.862 & 0.034 &  675 &  0.831 & 0.027 & 675 &  0.731 & 0.034 \\            
 700 &  0.809 & 0.025 &  700 &  0.801 & 0.028 & 700 &  0.727 & 0.027 \\            
 725 &  0.795 & 0.026 &  725 &  0.736 & 0.028 & 725 &  0.690 & 0.026 \\            
 750 &  0.723 & 0.028 &  750 &  0.683 & 0.025 & 750 &  0.641 & 0.024 \\            
 775 &  0.664 & 0.027 &  775 &  0.588 & 0.023 & 775 &  0.557 & 0.024 \\            
 800 &  0.594 & 0.026 &  800 &  0.541 & 0.024 & 800 &  0.537 & 0.030 \\            
 825 &  0.498 & 0.027 &  825 &  0.464 & 0.020 & 825 &  0.384 & 0.024 \\            
 850 &  0.439 & 0.022 &  850 &  0.349 & 0.025 & 850 &  0.315 & 0.022 \\            
 875 &  0.329 & 0.024 &  875 &  0.261 & 0.022 & 875 &  0.257 & 0.019 \\            
 900 &  0.171 & 0.024 &  900 &  0.166 & 0.019 & 900 &  0.135 & 0.018 \\            
 925 &  0.017 & 0.022 &  925 &  0.048 & 0.018 & 925 &  0.047 & 0.020 \\           
 950 & -0.176 & 0.022 &  950 & -0.110 & 0.018 & 950 & -0.133 & 0.021 \\         
 975 & -0.278 & 0.023 &  975 & -0.272 & 0.019 & 975 & -0.253 & 0.032 \\        
1000 & -0.397 & 0.023 & 1000 & -0.384 & 0.025 &     &        &       \\               
1025 & -0.494 & 0.029 & 1025 & -0.444 & 0.038 &     &        &       \\               
1050 & -0.651 & 0.042 & 1050 & -0.541 & 0.071 &     &        &       \\               
\hline                                                                          
\end{tabular}             
\caption{\it (continued).}                                                      
\end{table}                                                                   

\eject
{\Large\bf Figure captions} \\
\newcounter{fig}
\begin{list}   
{Figure \arabic{fig}.}
{\usecounter{fig}\setlength{\rightmargin}{\leftmargin}}
\item
{Experimental setup. D is the diamond target; 
 $K_{1-2}$ are collimators; $SM_{1-2}$ are 
 sweeping magnets; $H_2$ is the liquid hydrogen
 target; $Q$ is the Wilson quantameter; $M$ is
 the fast monitor; $L_{1-2}$ are quadrupole
 lenses; $PSM$ and $BM$ are bending magnets; $C
 _{1-2}$ is the thin converters; $Pb$ is the lead
 converter; $SF_{1-5},\,SB,\,BF_{1-6},\,BB_{1-6},
 \,S_{1-4},\,V,\,T$ are scintillation counters;
 $H_p,\,H_{xy}$ are hodoscopes; $\check{C}$ is
 the $\check{C}$erenkov counter; $PS-30$ is the 
 pair spectrometer; $MS$ is the magnetic
 spectrometer; $\check{C}_s$ is the
 $\check{C}$erenkov spectrometer.}
\label{f1}
\item
{Energy spectra of bremsstrahlung on 
 amorphous and crystal targets for different 
 coherent peak ($E_{\gamma}^{peak}$) and 
 electron ($E_e$) energies: spectrum with 
 amorphous target; coherent spectrum with 
 $E_e = 3\:GeV$ and $E_{\gamma}^{peak} = 
 600\:MeV$; coherent spectrum with $E_e = 
 3.5\:GeV$ and $E_{\gamma}^{peak} = 900\:MeV$.}
\label{f2}
\item
{Results of the test experiment with the 
 point-like target (solid histogram), 
 compared with the Monte Carlo predictions 
 (dashed histogram), for different proton 
 momenta ($P_p$) and target locations 
 ($X_{targ}$).}
\label{f3}
\item
{The peak region of the coherent bremsstrahlung 
 (shown with arrows), measured periodically 
 during the experiment, and used to check the 
 peak position.  The peak position is in the
 correct place (a); it is shifted to the side
 of lower energies (pair spectrometer channels)
 (b).}
\label{f4}
\item
{Time-of-flight measurements in MS.  The arrows
 indicate the cut positions selecting a proton
 candidate.  The left peak represents events
 with pion candidate in MS.}
\label{f5}
\item
{Start-stop measurements between two branches.}
\label{f6}
\item
{Energy dependence of $\pi^{\circ}$
 photoproduction asymmetry $\Sigma$ at $\theta^{*}
 = 90^{\circ}$ (a) and $120^{\circ}$ (b),
 respectively.  Experimental data are from
 Yerevan, present experiment (filled circles), 
 SLAC \protect\cite{zdar} (open triangles),  MIT
 \protect\cite{alest} (open circles), SLAC
 \protect\cite{knies} (open squares), Kharkov
 \protect\cite{ganen} (filled triangles), and
 Yerevan, previous measurements
 \protect\cite{yerold} (filled squares).  Solid
 (dash-dotted) curves give WI00 (FA00) results
 by GW \protect\cite{arndt} versus MAID2000
 results by the Mainz group \protect\cite{maid2}
 (dashed curves).}
\label{f7}
\item
{Energy dependence of $\pi^{\circ}$ 
 photoproduction asymmetry $\Sigma$ at 
 $\theta^{*} = 100^{\circ}$ (a) and $110^{\circ}$
 (b), respectively.  Notation is the same as is 
 in Fig.~\ref{f7}.}
\label{f8}
\item
{Angular dependence of 
 $\pi^{\circ}$ photoproduction asymmetry $\Sigma$
 at $E_{\gamma} = 700\,MeV$ (a), $E_{\gamma} =
 750\,MeV$ (b), and $E_{\gamma} = 800\,MeV$ (c),
 respectively.  Notation is the same as is in
 Fig.~\ref{f7}.}
\label{f9}
\item
{Angular dependence of $\pi^{\circ}$ (a) and
 $\pi^+$ (b) photoproduction asymmetry $\Sigma$ 
 at $E_{\gamma} = 950\,MeV$.  $\pi^{\circ}$
 data are present measurements.  $\pi^+$ data
 are from GRAAL \protect\cite{GRAAL} (filled
 asterisk) and DNPL \protect\cite{DNPL} (open
 diamond). Plotted are fits FA00 (dash-dotted),
 FA00 with a modified $E^{1/2}_{0+}$ multipole
 (see text) (solid), and the MAID2000
 result\protect\cite{maid2} (dashed).}
\label{f10}
\end{list}
\eject
\begin{figure}[tb!]
\centerline{\psfig{file=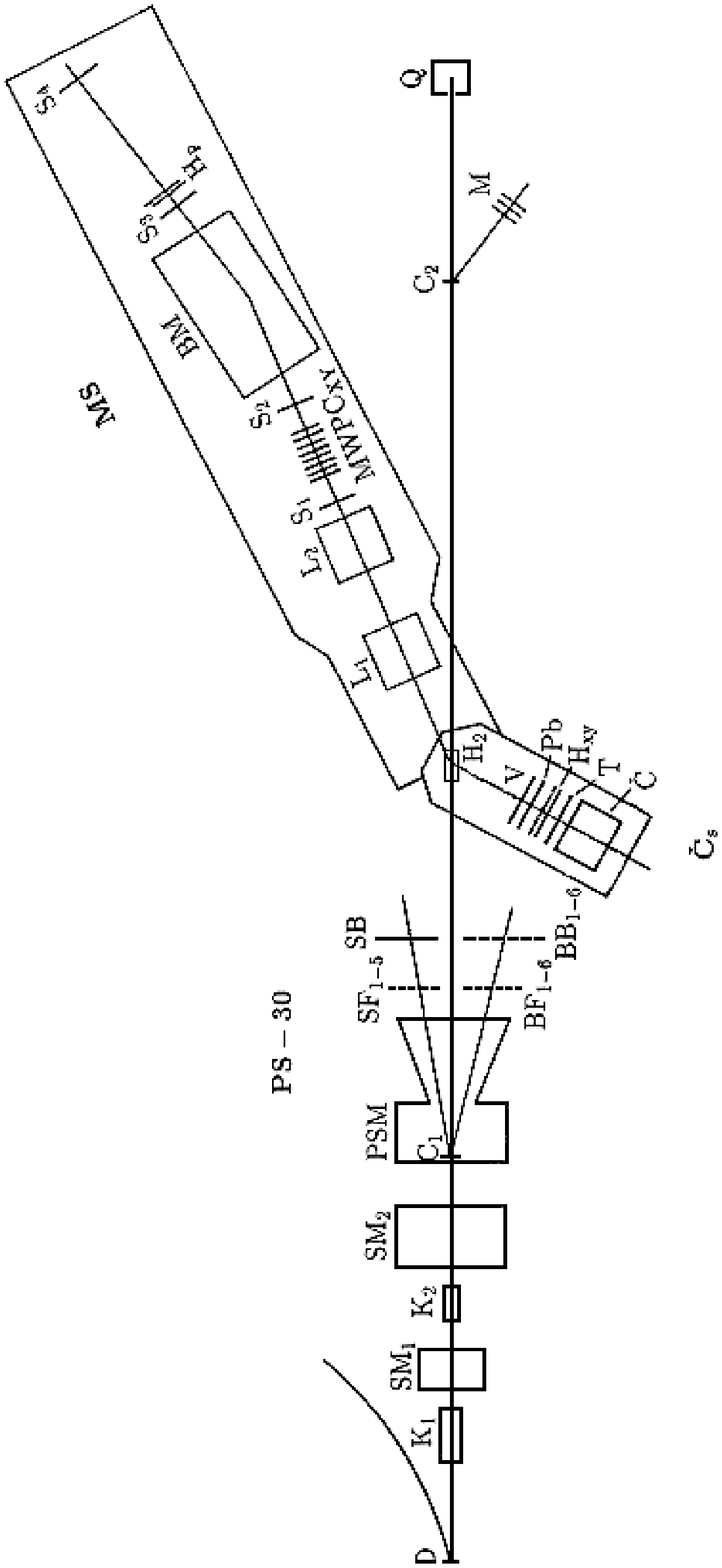,width=.6\textwidth,clip=,silent=,angle=0}}
\caption[fig1a]{\label{fig1}}
\end{figure}
\begin{figure}[tb!]
\centerline{\psfig{file=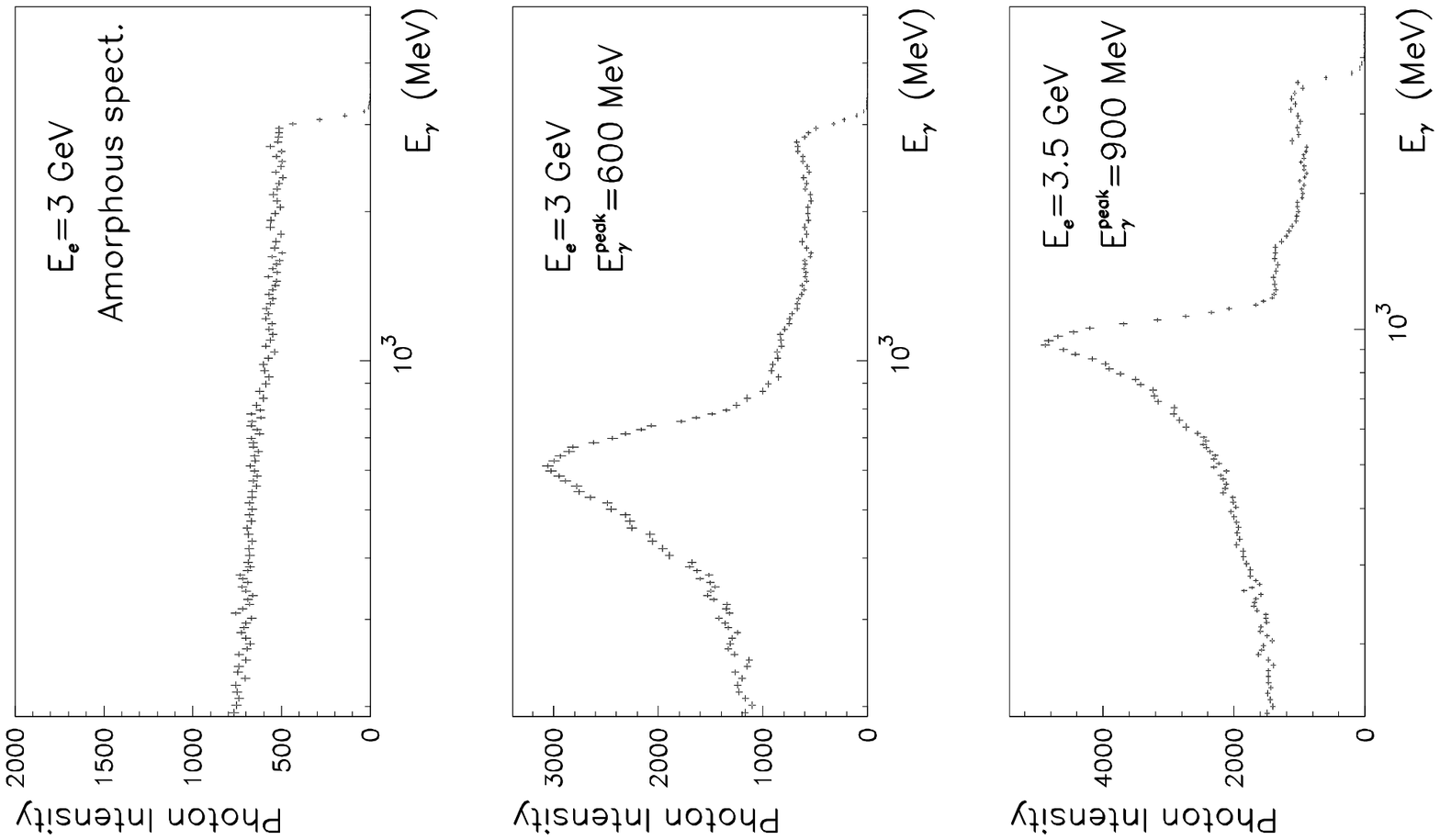,width=.6\textwidth,clip=,silent=,angle=90}}
\caption[fig1a]{\label{fig2}}
\end{figure}
\begin{figure}[tb!]
\centerline{\psfig{file=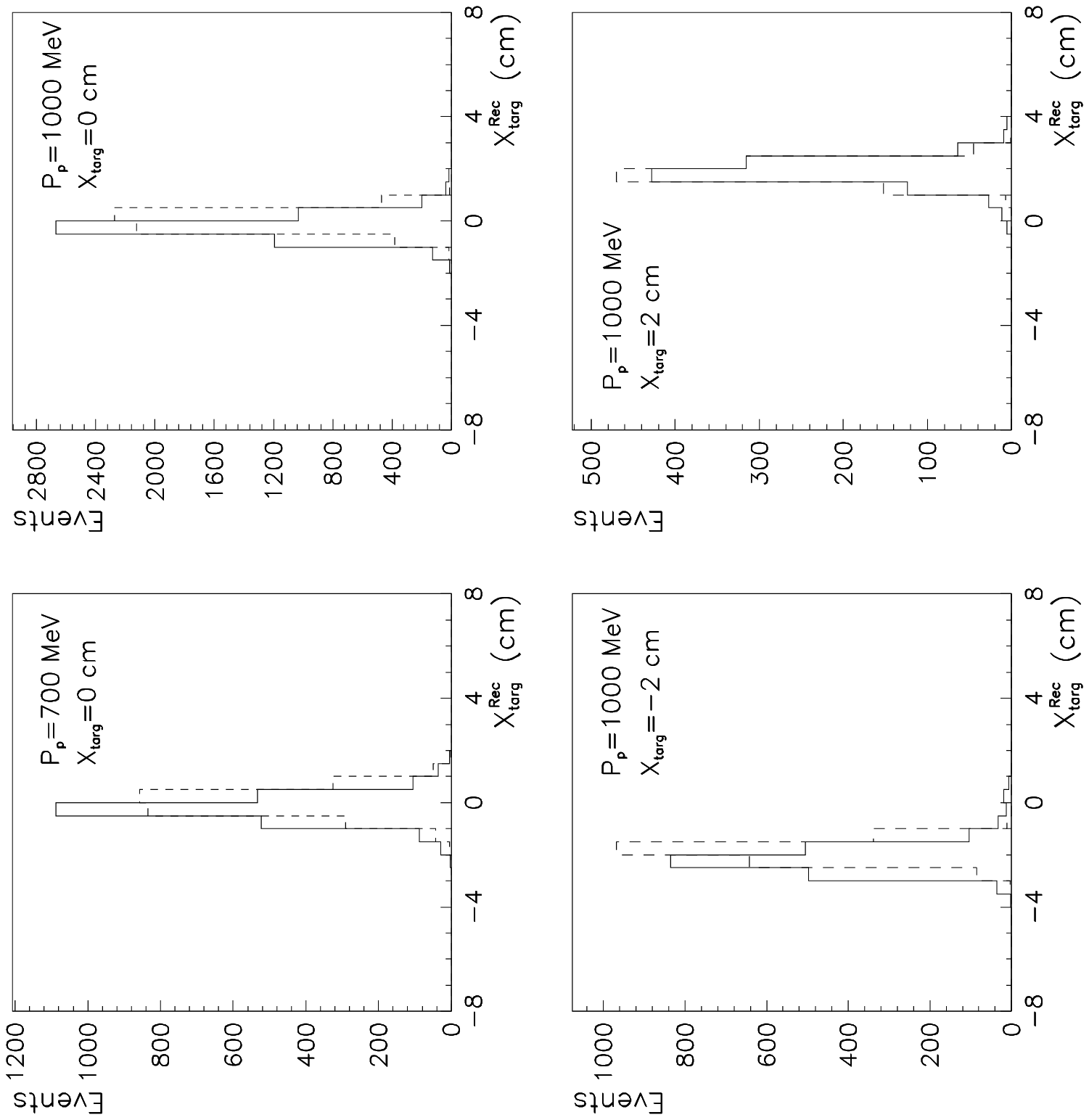,width=.6\textwidth,clip=,silent=,angle=90}}
\caption[fig1a]{\label{fig3}}
\end{figure}
\begin{figure}[tb!]
\centerline{\psfig{file=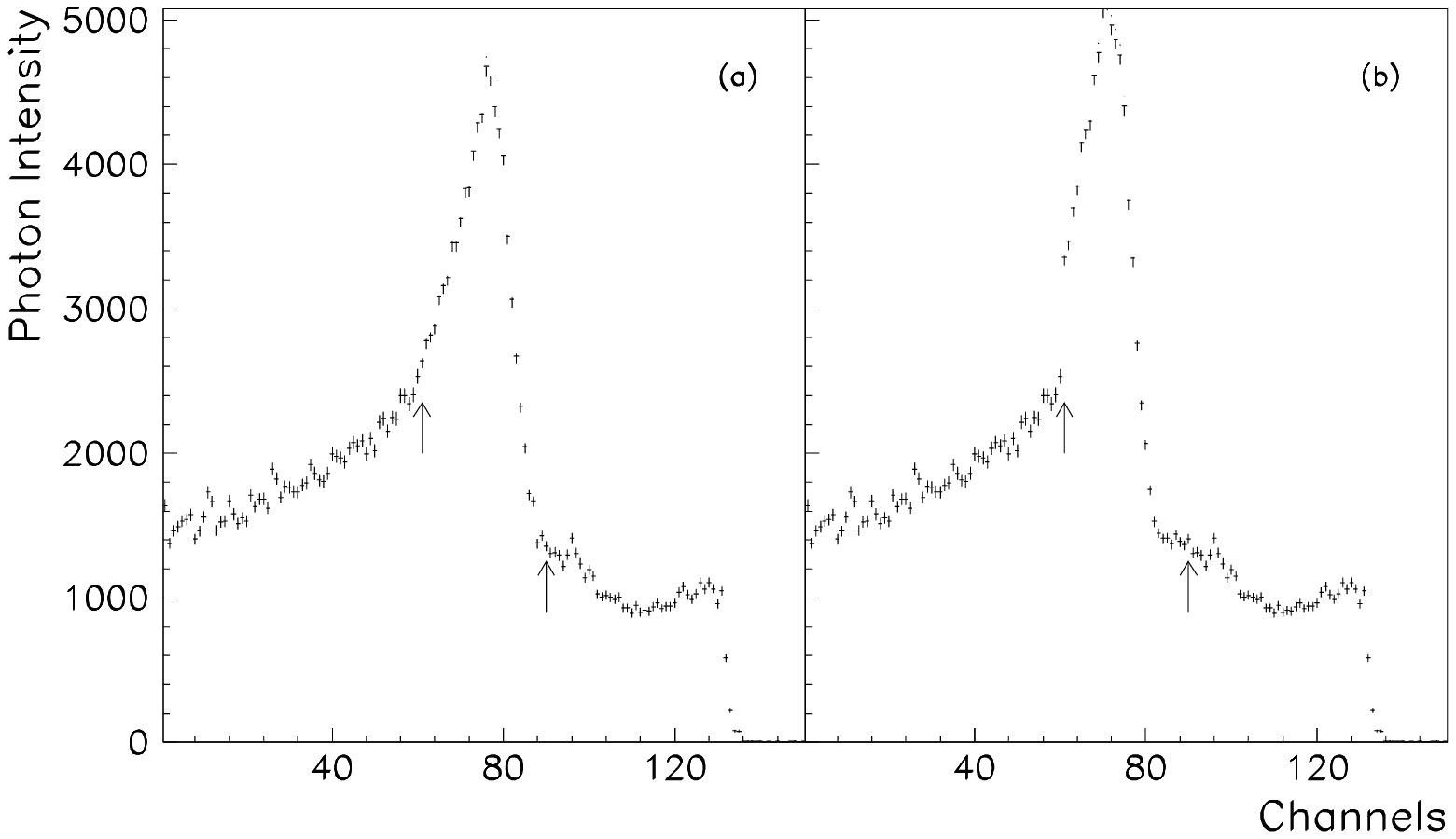,width=.7\textwidth,clip=,silent=,angle=0}}
\caption[fig1a]{\label{fig4}}
\end{figure}
\begin{figure}[tb!]
\centerline{\psfig{file=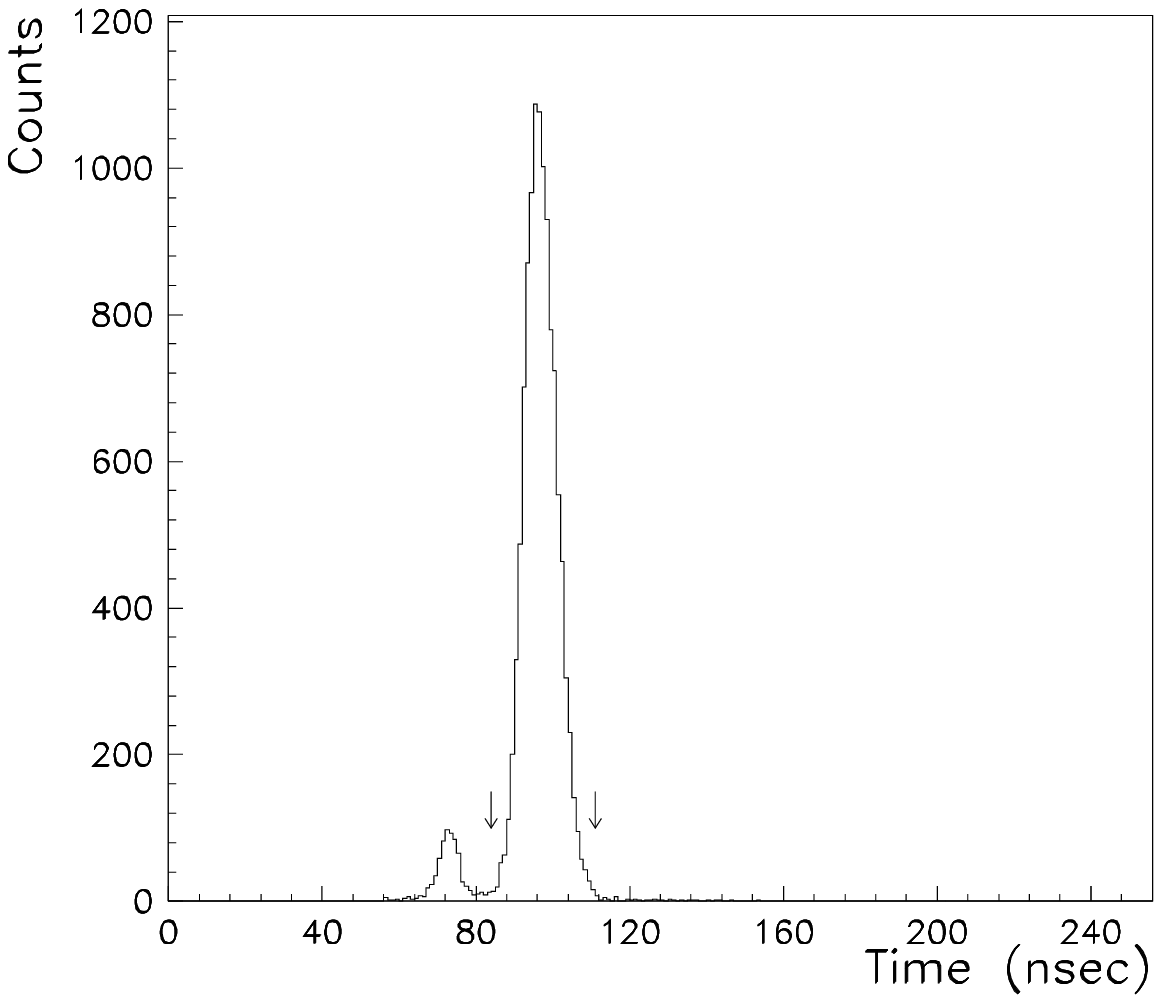,width=.5\textwidth,clip=,silent=,angle=0}}
\caption[fig1a]{\label{fig5}}
\end{figure}
\begin{figure}[tb!]
\centerline{\psfig{file=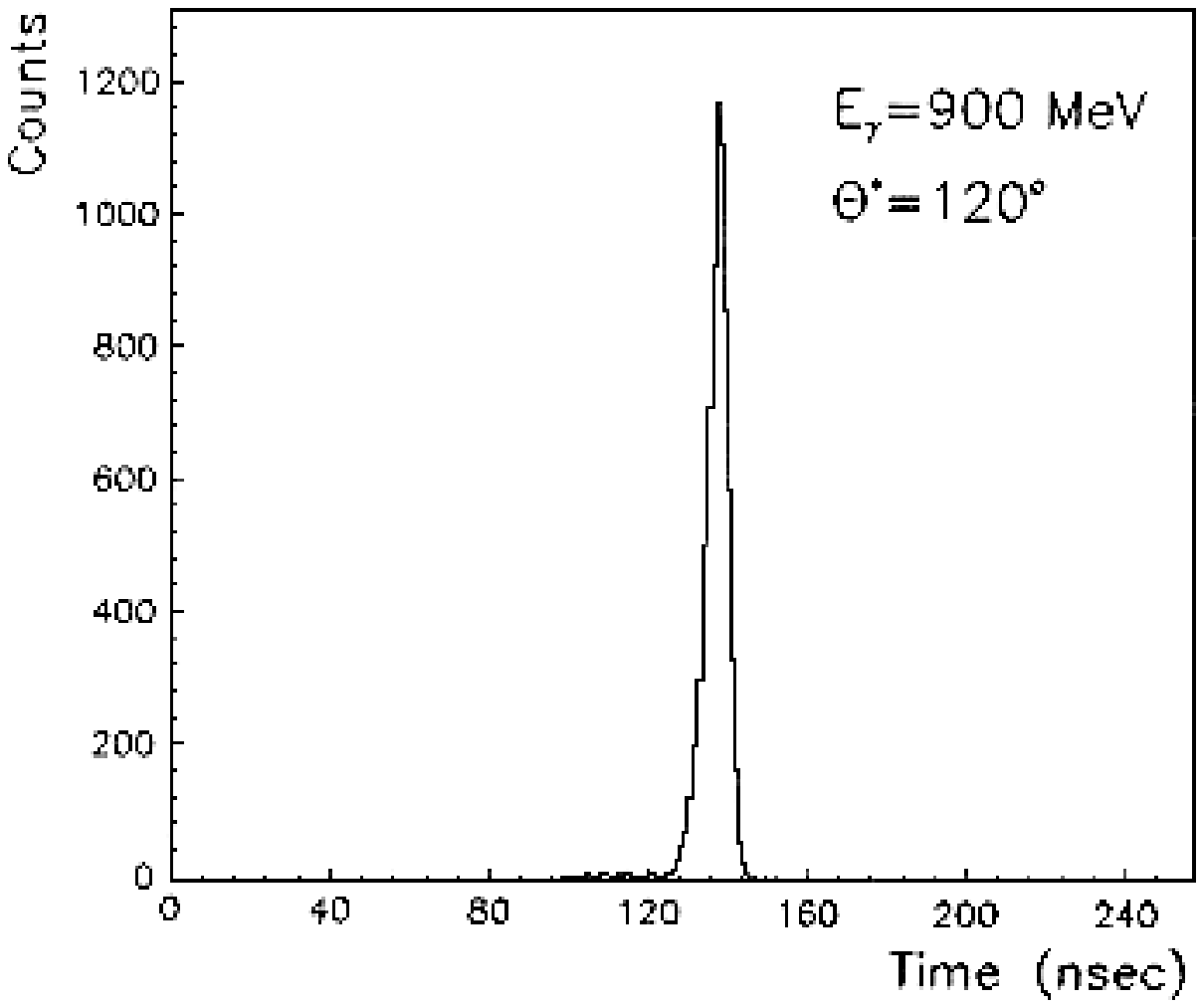,width=.4\textwidth,clip=,silent=,angle=0}}
\caption[fig1a]{\label{fig6}}
\end{figure}
\begin{figure}[tb!]
\centerline{\psfig{file=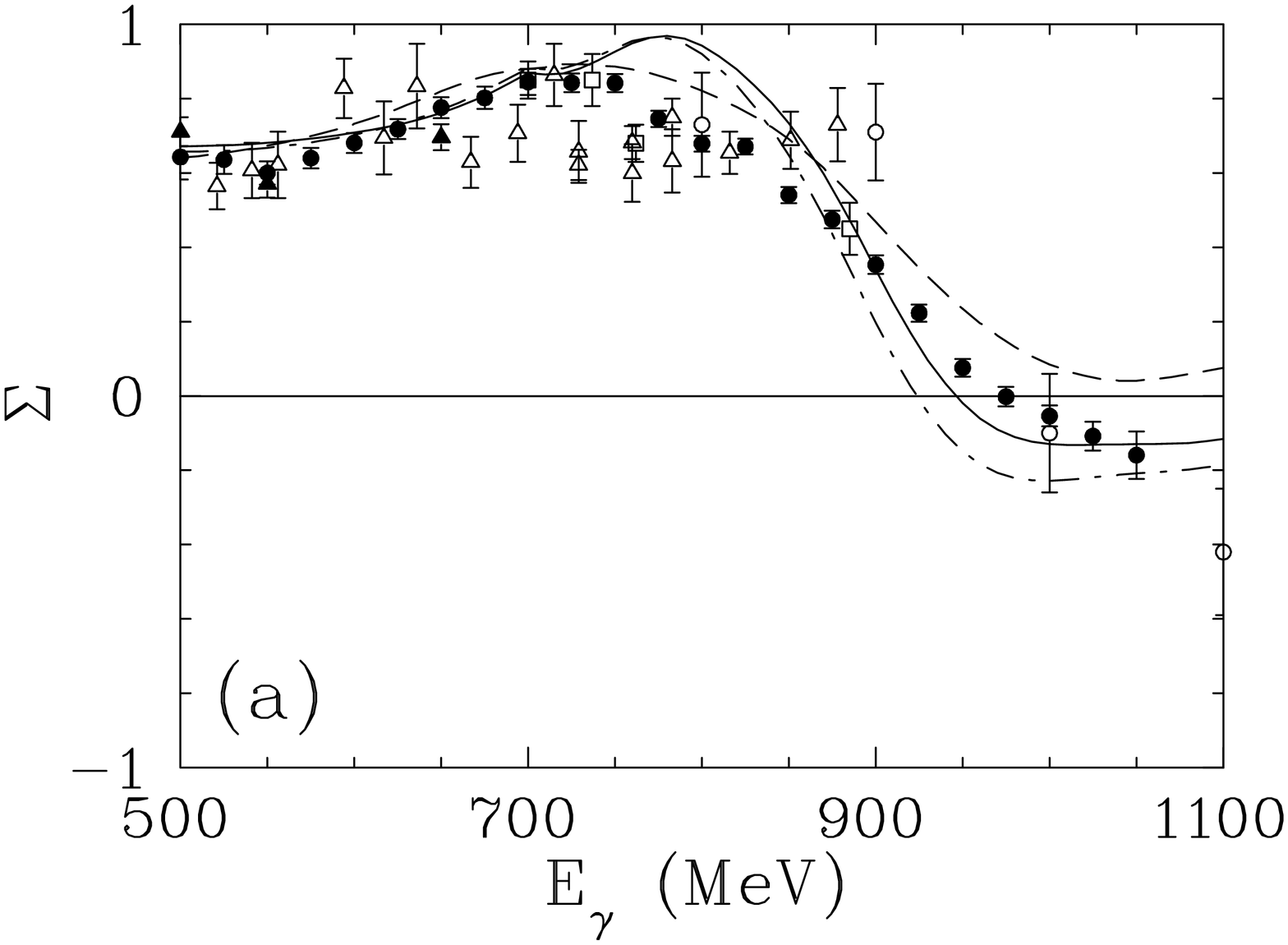,width=.8\textwidth,clip=,silent=,angle=90}}
\caption[fig1a]{\label{fig7a}}
\end{figure}
\begin{figure}[tb!]
\centerline{\psfig{file=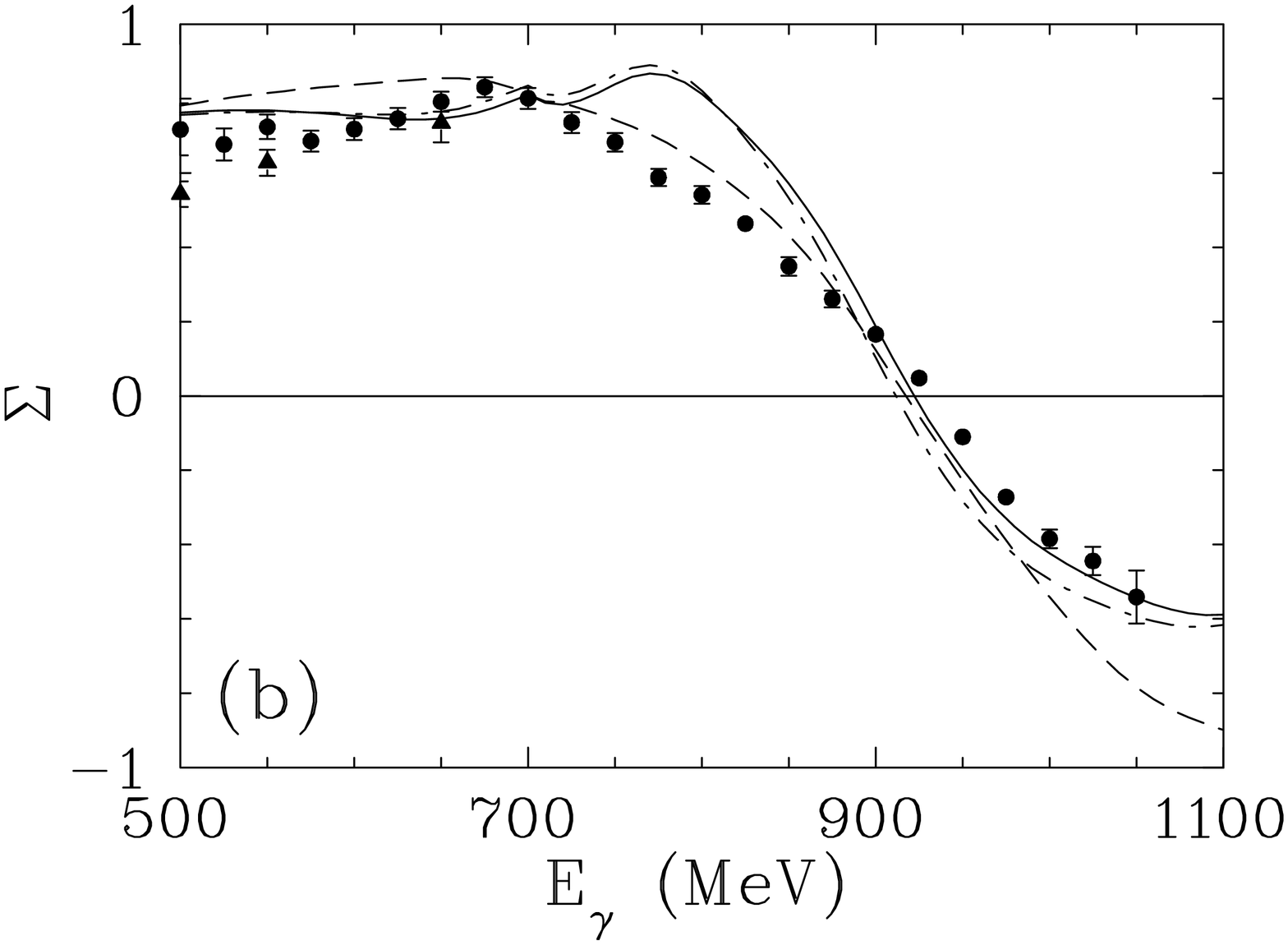,width=.8\textwidth,clip=,silent=,angle=90}}
\caption[fig1a]{\label{fig7b}}
\end{figure}
\begin{figure}[tb!]
\centerline{\psfig{file=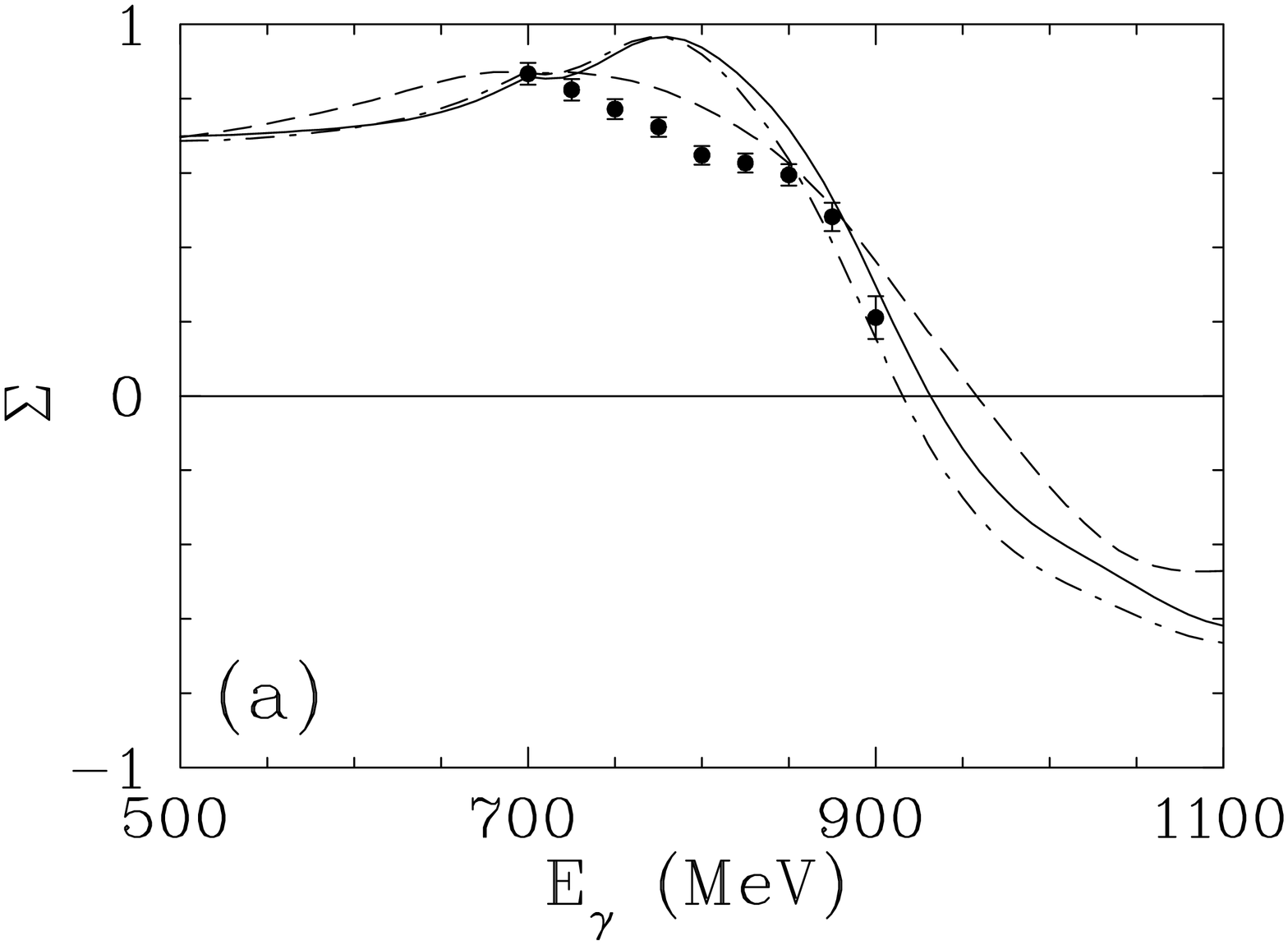,width=.8\textwidth,clip=,silent=,angle=90}}
\caption[fig1a]{\label{fig8a}}
\end{figure}
\begin{figure}[tb!]
\centerline{\psfig{file=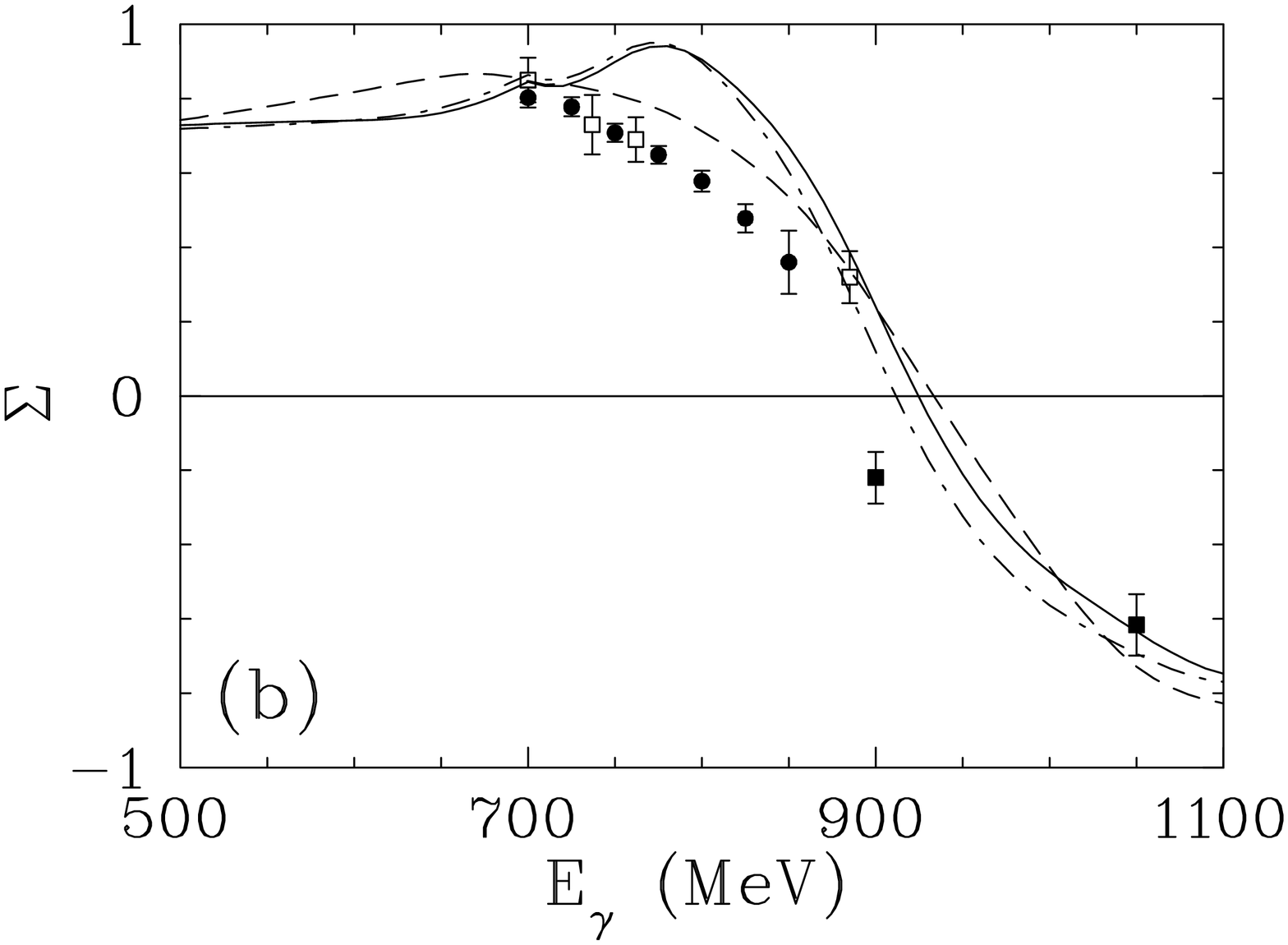,width=.8\textwidth,clip=,silent=,angle=90}}
\caption[fig1a]{\label{fig8b}}
\end{figure}
\begin{figure}[tb!]
\centerline{\psfig{file=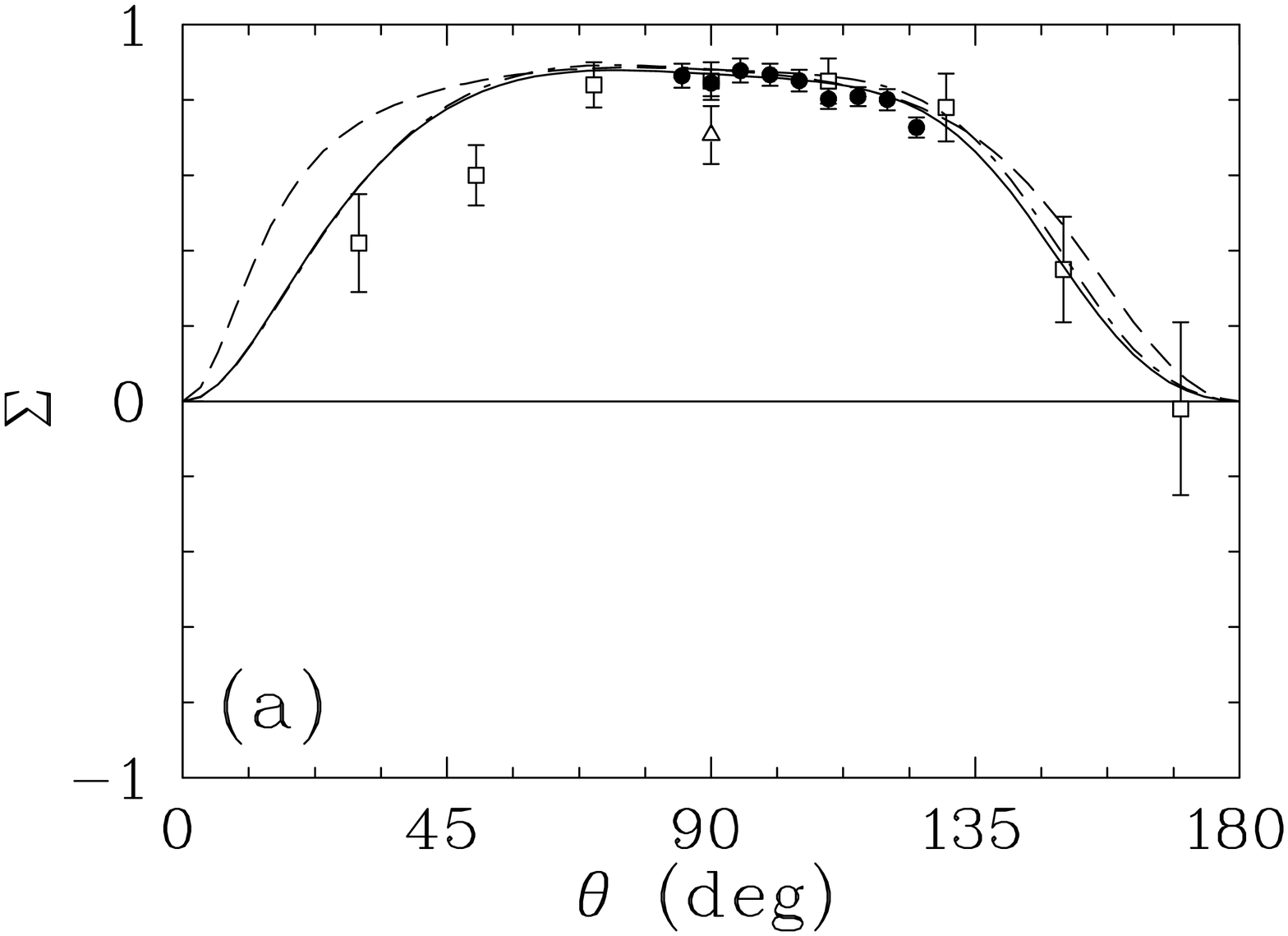,width=.8\textwidth,clip=,silent=,angle=90}}
\caption[fig1a]{\label{fig9a}}
\end{figure}
\begin{figure}[tb!]
\centerline{\psfig{file=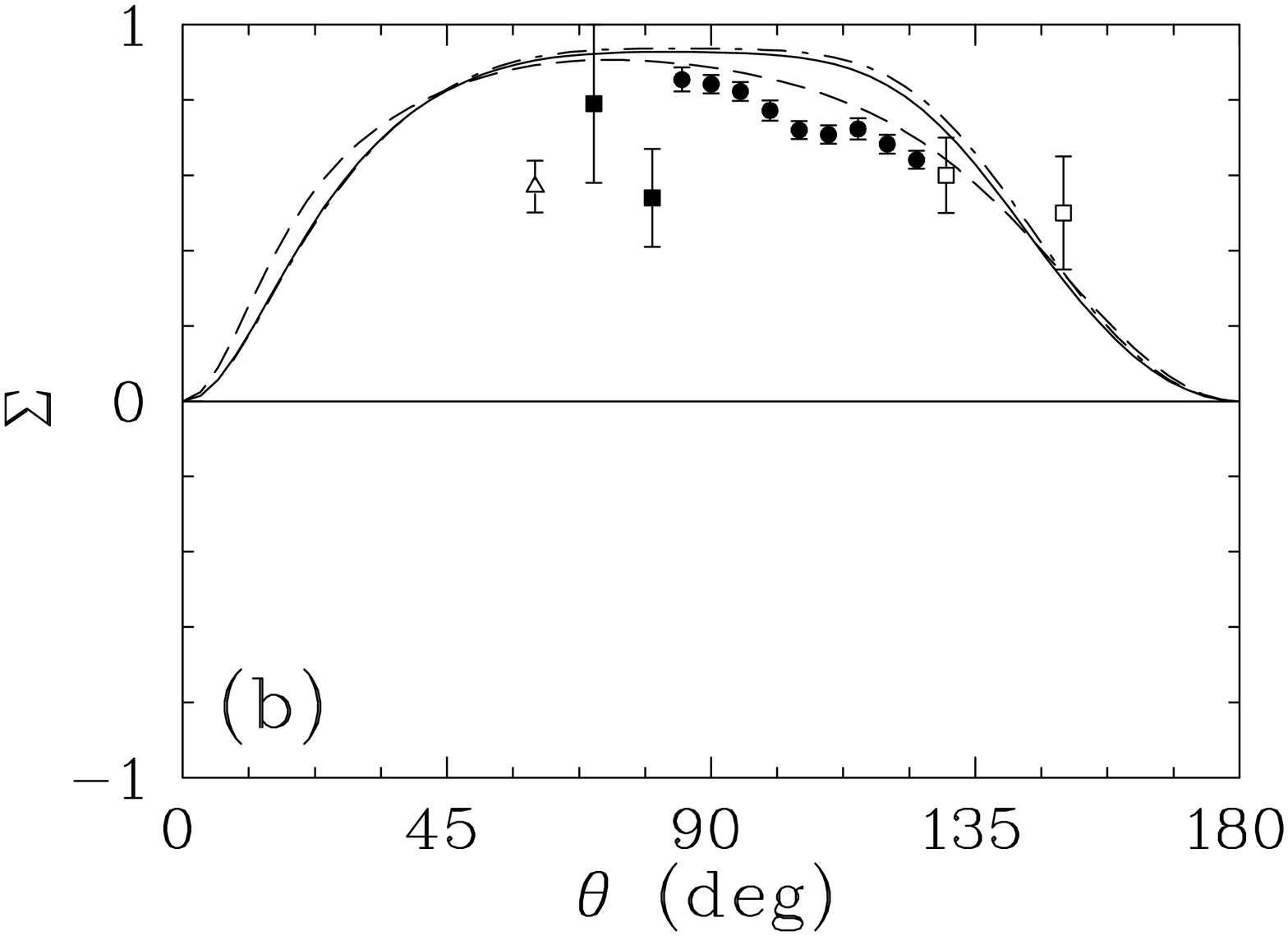,width=.8\textwidth,clip=,silent=,angle=90}}
\caption[fig1a]{\label{fig9b}}
\end{figure}
\begin{figure}[tb!]
\centerline{\psfig{file=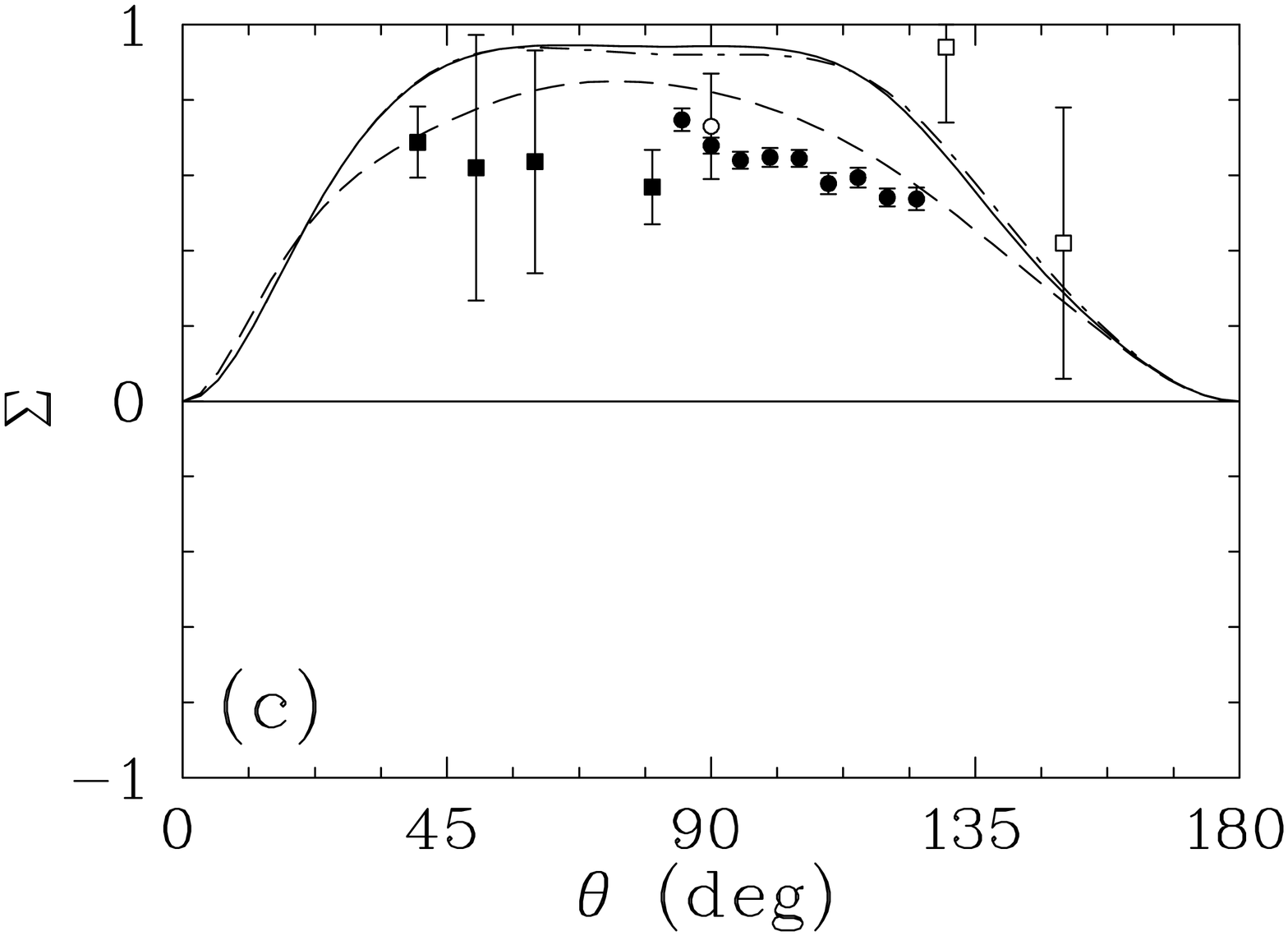,width=.8\textwidth,clip=,silent=,angle=90}}
\caption[fig1a]{\label{fig9c}}
\end{figure}
\begin{figure}[tb!]
\centerline{\psfig{file=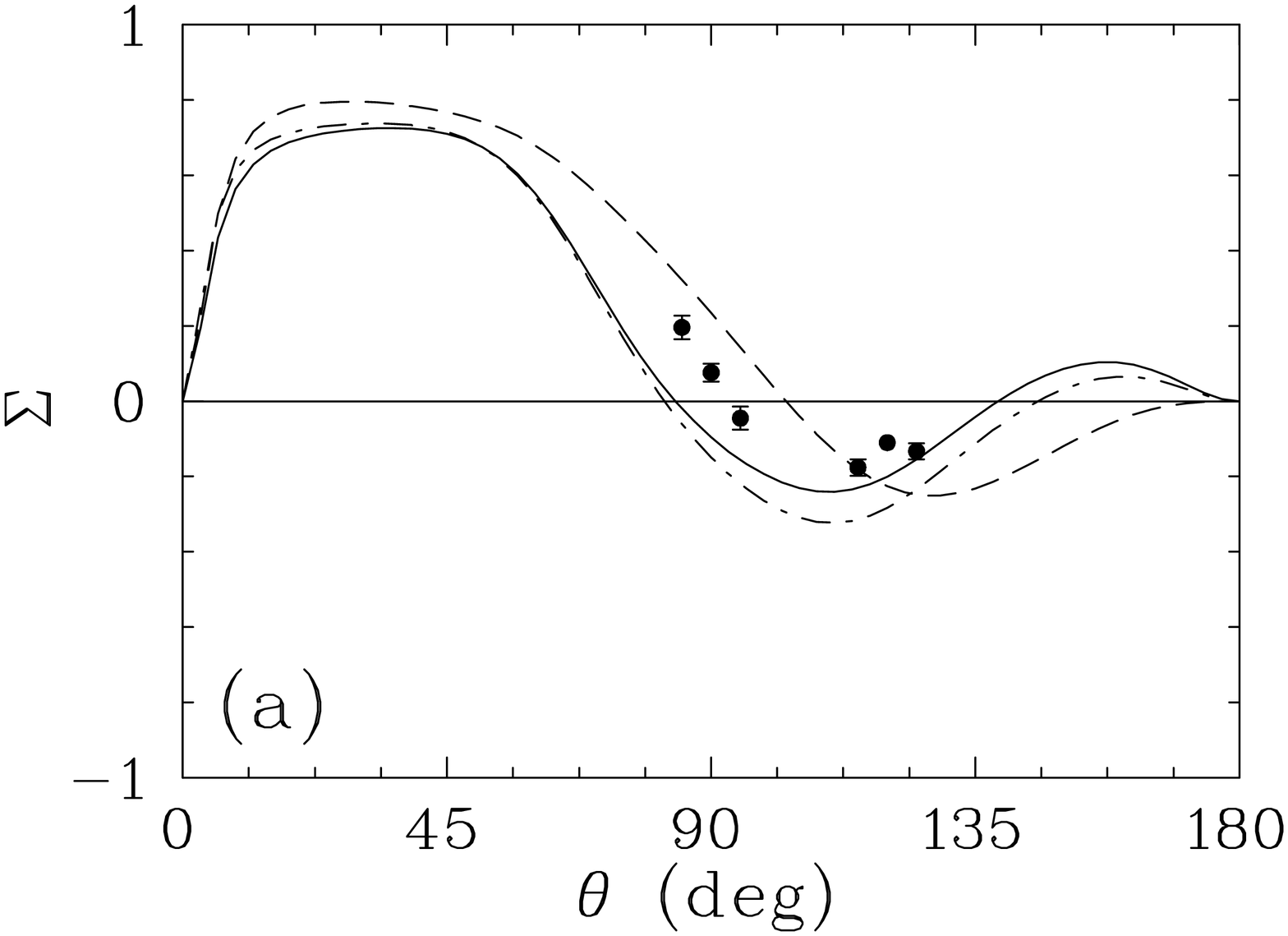,width=.8\textwidth,clip=,silent=,angle=90}}
\caption[fig1a]{\label{fig10a}}
\end{figure}
\begin{figure}[tb!]
\centerline{\psfig{file=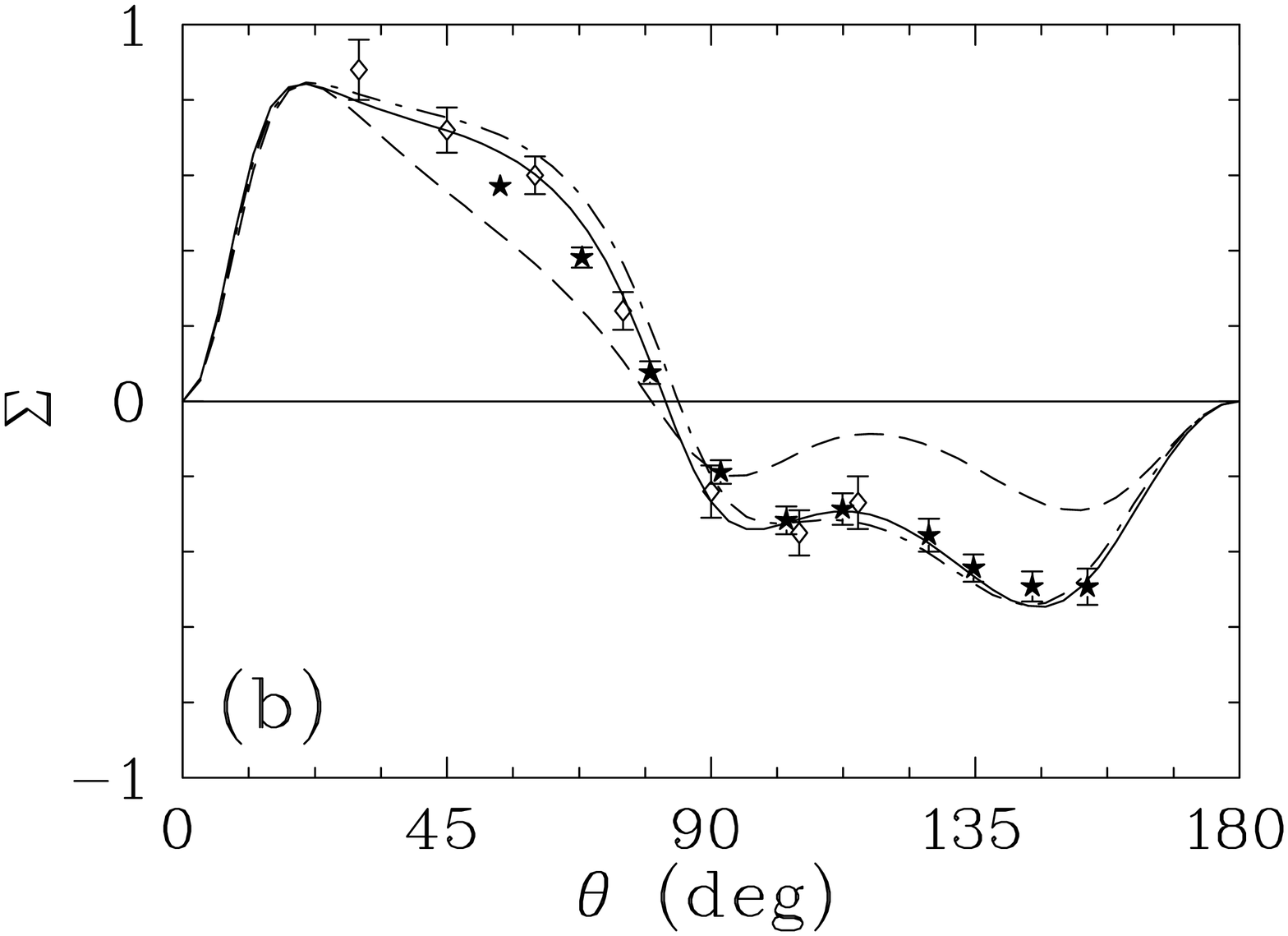,width=.8\textwidth,clip=,silent=,angle=90}}
\caption[fig1a]{\label{fig10b}}
\end{figure}


\begin{thebibliography}{99}
\bibitem{pdg}
D. E. Groom \textit{et al.}, Eur. Phys. J. 
\textbf{C15}, 1 (2000).

\bibitem{e2m1}
R. M. Davidson, N. C. Mukhopadhyay, M. S. Pierce, 
R. A. Arndt, I. I. Strakovsky, and R. L. Workman, 
Phys. Rev. C \textbf{59}, 1059 (1999). 

\bibitem{etaphoto1}
L. Tiator, D. Drechsel, G. Kn\"ochlein, and C. 
Bennhold, Phys. Rev. C \textbf{60}, 035210 
(1999). 

\bibitem{etaphoto2}
R. L. Workman, R. A. Arndt, and I. I. Strakovsky, 
Phys. Rev. C \textbf{62}, 048201 (2000).

\bibitem{kaonphoto}
T. Mart and C. Bennhold, Phys. Rev. C \textbf{61}, 
012201 (1999).

\bibitem{GRAAL}
J. Ajaka \textit{et al.}, Phys. Lett. 
\textbf{B475}, 372 (2000).

\bibitem{vartap}
A. Vartapetian. Ph.D. Thesis, Hamburg University, 1994; \\
http://home.cern.ch/$\sim$vartap/phd/phd.ps.gz

\bibitem{adam}
F. V. Adamian \textit{et al.}, J. Phys. \textbf{G19}, 
L139 (1993).

\bibitem{pi0}
F. V. Adamian \textit{et al.}, Preprint 
YERPHI--1488(5)--97, Yerevan.

\bibitem{mainz}
D. Drechsel, O. Hanstein, S. S. Kamalov, and L. Tiator, 
Nucl. Phys. \textbf{A645}, 145 (1999).

\bibitem{adam2}
F. V. Adamian \textit{et al.}, J. Phys. \textbf{G17}, 
1189 (1991).

\bibitem{hrach}
H. H. Akopian \textit{et al.}, Preprint 
YERPHI--908(59)--86, Yerevan; \\ 
http://home.cern.ch/$\sim$vartap/scanned\_preprints/908\_86.html

\bibitem{zdar}
R. Zdarko and E. Dally, Nuovo Cim. \textbf{A10}, 
10 (1972).

\bibitem{alest}
J. Alspector \textit{et al.}, Phys. Rev. Lett. 
\textbf{28}, 1403 (1972).

\bibitem{knies}
G. Knies \textit{et al.}, Phys. Rev. D \textbf{10}, 
2778 (1974).

\bibitem{ganen}
V. Ganenko \textit{et al.}, Phys. At. Nucl. (former 
Sov. J. Nucl. Phys.) \textbf{23}, 162 (1976); A. A. 
Belyaev \textit{et al.}, Nucl. Phys. \textbf{B213}, 
201 (1983) and references therein.

\bibitem{yerold}
R. O. Avakyan \textit{et al.}, Phys. At. Nucl. 
(former Sov. J. Nucl. Phys.) \textbf{40}, 588 
(1984) and references therein.

\bibitem{maid2}
L. Tiator {\em {et al.}}, in preparation.

\bibitem{arndt}
R. A. Arndt \textit{et al.}, Phys. Rev. C 
\textbf{53}, 430 (1996); R. A. Arndt \textit{et al.}, 
in preparation.

\bibitem{cebaf1}
W. J. Briscoe \textit{et al.}, TJNAF/CEBAF experiment 
proposal E--94--103, 1994.

\bibitem{barb}
I. M. Barbour \textit{et al.}, Nucl. Phys. 
\textbf{B141}, 253 (1978).

\bibitem{tiator}
L. Tiator, private communications, 2000.

\bibitem{cebaf2}
D. F. Geesman \textit{et al.}, TJNAF/CEBAF 
experiment proposal E--94--012, 1994, \\ D. 
Jenkins, TJNAF/CEBAF Letter of Intent 
LOI--96--001, 1996. 

\bibitem{DNPL}
P. J. Bussey \textit{et al.}, Nucl. Phys. 
\textbf{B154}, 205 (1979).

\bibitem{graalnew}
V. Kouznetsov \textit{et al.}, to be published 
in Proceedings of IX Moscow International 
Seminar on Electromagnetic Interactions off 
Nuclei at Low and Medium Energies, Institute 
for Nuclear Research, Moscow, Russia, 21--23 
Sept. 2000.

\end{thebibliography}
\end{document}